\documentclass{sig-alternate}
\newfont{\mycrnotice}{ptmr8t at 7pt}
\newfont{\myconfname}{ptmri8t at 7pt}
\let\crnotice\mycrnotice%
\let\confname\myconfname%

\usepackage[german,american]{babel}
\usepackage[T1]{fontenc}
\usepackage[utf8]{inputenc}
\usepackage{times}
\usepackage{textcomp}
\usepackage{graphicx}
\usepackage{amssymb}
\usepackage{microtype} % for linebreaks in texttt
\usepackage[hidelinks]{hyperref}
\usepackage{url}
\renewcommand{\tt}{\fontencoding{OT1}\fontfamily{cmtt}\selectfont}

\usepackage[usenames,dvipsnames]{color}
\usepackage{booktabs}
\usepackage[numbers,sectionbib]{natbib}%[round]
\makeatletter
\def\@copyrightspace{\relax}
\makeatother

\begin{document}

\title{Overview of the Triple Scoring Task at the WSDM Cup 2017}

\numberofauthors{3}
\author{
\alignauthor
Hannah Bast\\
\affaddr{University of Freiburg}\\
\affaddr{bast@cs.uni-freiburg}\\
\alignauthor
Bj\"{o}rn Buchhold\\
\affaddr{University of Freiburg}\\
\affaddr{buchhold@cs.uni-freiburg}\\
\alignauthor
Elmar Haussmann\\
\affaddr{University of Freiburg}\\
\affaddr{haussmann@cs.uni-freiburg}
}

\maketitle

\begin{abstract}
This paper provides an overview of the triple scoring task at the WSDM Cup 2017, including a description of the task and the dataset,
an overview of the participating teams and their results, and a brief account of the methods employed.
In a nutshell, the task was to compute relevance scores for knowledge-base triples from relations, where such scores make sense.
Due to the way the ground truth was constructed, scores were required to be integers from the range 0..7.
For example, reasonable scores for the triples \emph{Tim Burton profession Director} and \emph{Tim Burton profession Actor} would be $7$ and $2$, respectively, because Tim Burton is well-known as a director, but he acted only in a few lesser known movies.

The triple scoring task attracted considerable interest, with 52 initial registrations and 21 teams who submitted a valid run before the deadline.
The winning team achieved an accuracy of 87\%, that is, for that fraction of the triples from the test set (which was revealed only after the deadline) the difference to the score from the ground truth was at most 2. The best result for the average difference from the test set scores was $1.50$.
\end{abstract}

%%%%%%%%%%%%%%%%%%%%%%%%%%%%%%%%%%%%%%%%%%%%%%%%%%%%%%%%%%%%%%%%%%%%%%%%
\section{Introduction}

Knowledge bases allow queries that express the search intent precisely. For example, we can easily formulate a query that gives us precisely a list of all \emph{American actors} in a knowledge base. Note the fundamental difference to full-text search, where keyword queries are only approximations of the actual search intent, and thus result lists are typically a mix of relevant and irrelevant hits.

But even for result sets containing only relevant items, a ranking of the contained items is often desirable. One reason is similar as in full-text search: when the result set is very large, we cannot look at all items and thus want the most ``interesting'' items first.
But even for small result sets, it is useful to show the inherent order of the items in case there is one. 
We give two examples.
The numbers refer to a sanitized dump of Freebase from June 29, 2014; see \cite{DBLP:conf/www/BastBBH14}.

\medskip\noindent
{\bf Example 1 (American actors):} Consider the query that returns all entities that have \emph{Actor} as their profession and \emph{American} as their nationality. On the mentioned version of Freebase, this query has 64,757 matches. A straightforward ranking would be by popularity, as measured, e.g., by counting the number of occurrences of each entity in a reference text corpus. Doing that, the top-5 results for our query look as follows (the first result is G.~W.~Bush):\\[1mm]
\noindent
\emph{George Bush},\emph{Hillary Clinton},\emph{Tim Burton},\emph{Lady Gaga},\emph{Johnny Depp}\\[1mm]
All five of these are indeed listed as actors in Freebase.
This is correct in the sense that each of them appeared in a number of movies, and be it only in documentary movies as themselves or in short cameo roles.
However, Bush and Clinton are known as politicians, Burton is known as a film director, and Lady Gaga as a musician.
Only Johnny Depp, number five in the list above, is primarily an actor.
He should be ranked before the other four.

\medskip\noindent
{\bf Example 2 (professions of a single person):} Consider all professions by Arnold Schwarzenegger.
Freebase lists 10 entries:\\[1mm]
\noindent
\emph{Actor},
\emph{Athlete},
\emph{Bodybuilder},
\emph{Businessperson},
\emph{Entrepreneur},
\emph{Film Producer},
\emph{Investor},
\emph{Politician},
\emph{Television Director},
\emph{Writer}\\[1mm]
Again, all of them are correct in a sense.
For this query, ranking by ``popularity'' (of the professions) makes even less sense than for the query from Example~1.
Rather, we would like to have the ``main'' professions of that particular person at the top.
For Arnold Schwarzenegger that would be: \emph{Actor}, \emph{Politician}, \emph{Bodybuilder}.
Note how we have an ill-defined task here: it is debatable whether Arnold Schwarzenegger is more of an actor or more of a politician.
But he is certainly more of an actor than a writer.

\subsection{Task Definition}\label{sec:definition}

\enlargethispage{\baselineskip}
The task is to compute relevance scores for triples from type-like relations.
The following definition is adapted from \cite{DBLP:conf/sigir/BastBH15}:
% (where scores are normalized to the range [0,1]):\\[1mm]

\medskip\noindent
{\bf Definition:}
Given a list of triples from two type-like relations (profession and nationality), for each triple compute an integer score from $0..7$ that measures the degree to which the subject belongs to the respective type (expressed by the predicate and object).

\medskip
Here are four example scores, some of which are related to the example queries above. The numbers in parentheses are the respective scores normalized to the more intuitive range $[0, 1]$ (that is, divided by 7) and rounded to two digits after the dot.
\def\score#1{{\small (#1)}}
\begin{center}
\vspace{-0.5ex}
\begin{tabular}{@{\hspace{2mm}}l@{\hspace{6mm}}l@{\hspace{6mm}}l@{\hspace{6mm}}l@{\hspace{4mm}}l@{\hspace{2mm}}}
\emph{Tim Burton}    & \emph{profession}  & \emph{Director}      & $7$ & \score{1.00} \\
\emph{Tim Burton}    & \emph{profession}  & \emph{Actor}         & $3$ & \score{0.43} \\
\emph{Johnny Depp}   & \emph{profession}  & \emph{Actor}         & $7$ & \score{1.00} \\
\emph{Roger Federer} & \emph{nationality} & \emph{Swiss}         & $7$ & \score{1.00} \\
\emph{Roger Federer} & \emph{nationality} & \emph{South African} & $1$ & \score{0.14} \\
\end{tabular}
\vspace{-2mm}
\end{center}

\noindent
An alternative way of expressing this notion of degree is:
How ``surprised'' would we be to see, for example, Roger Federer in a list of South Africans.
This information is correct, yet most people would be very surprised, hence the low relevance score above.
The ``surprised'' formulation is also used in the crowdsourcing task which we designed to acquire 
human judgments for the ground truth used in our evaluation.

\subsection{Datasets and Evaluation}\label{sec:data}

Participants were provided a knowledge base in the form of 818,023 triples from two Freebase relations: \emph{profession} and \emph{nationality}.
Overall, these triples contain 376,214 different subjects, 200 different professions, and 100 different nationalities.
% HB: in the first version of the paper and on the website it says 385,426. I checked (21-12-2017) that this is a strict superset.

We constructed a ground truth for 1,387 of these triples (1,028 profession, 359 nationality).
For each triple, we obtained 7 binary relevance judgments from a carefully implemented and controlled crowdsourcing task, as described in \cite{DBLP:conf/sigir/BastBH15}.
This gives a total of 9,709 relevance judgments.
For each triple, the sum of the binary relevance judgments yields the score, which can be any integer in the range 0..7.

About half of this ground truth (677 triples) was made available to the participants as training data.
This was useful for understanding the task and the notion of ``degree'' in the definition above.
However, the learning task was still inherently unsupervised, because the training data covers only a subset of all professions and nationalities.
Participants were allowed to use arbitrary external data for unsupervised learning.
For convenience, we provided 33,159,353 sentences from Wikipedia for most (all but 407) subjects from the knowledge base.
For each subject from the ground truth, there was at least one sentence (and usually many more) with that subject annotated.
Not surprisingly, all participating teams made use of this data.

%are summed used to compute two scores:
%a discrete integer score between~0 and~7 (by simply taking the number of primary relevance votes), and a normalized score between~0 and~1 (by dividing the integer score by~7).
%This dataset will be used as a ground truth for testing only.
%Participants are supposed to compute the scores in an unsupervised manner.

%\medskip
The submissions were evaluated on a test set consisting of 710 triples.
These triples were known to be a subset of the 818,023 triples from the provided knowledge base.
However, it was (of course) not known before the submission deadline, which subset this was.

Three quality measures were applied to measure the quality of participating systems with respect to our ground truth:\\[1mm]
\noindent \emph{Accuracy (ACC):} the percentage of triples for which the score (an integer from the range 0..7) differs by at most~2 (in either direction) from the score in the ground truth.\\[1mm]
\noindent \emph{Average score difference (ASD):} the average (over all triples in the ground truth) of the absolute difference of the score computed by the participating system and the score from the ground truth.\\[1mm]
\noindent \emph{Kendall's Tau (TAU):} a rank-based measure which compares the ranking of all the professions (or nationalities) of a person with the ranking computed from the ground truth scores.
The handling of items with equal score is described in \cite[Section~5.1]{DBLP:conf/sigir/BastBH15} and under the link at the end of this section.\\[-2mm]

%Because scores in the ground truth are discrete, items often share a rank.
%Such a ranking with ties constitutes a \emph{partial ranking}. 
%To compare partial rankings we use an adapted version of Kendall's $\tau$:
%$\tau_p = \frac{1}{Z}(n_d + p \cdot n_t)$,
%where $n_d$ is the number of \emph{discordant} (inverted) pairs,
%$n_t$ is the number of pairs that are tied in the ground truth but not in the predicted ranking or vice versa,
%$p$ is a penalization factor for these pairs which we set to $0.5$,
%and the normalization factor $Z$ (the number of ordered pairs plus $p$ times the number of tied pairs in the ground truth).

The winners of the competition (there was a money prize for the top 3) were determined according to the ranking with respect to the ACC measure. If a team submitted more than one valid run, the last valid run submitted before the deadline was considered.

It should be noted that the ACC measure can only increase (and never decrease) when all scores 0 and 1 are rounded up to 2, and all scores 6 and 7 are rounded down to 5.
We call this the \emph{2-5-trick} in the following.
For winning the competition, it was advantageous to apply this trick, but (as we will see in Section \ref{sec:results}) the other measures suffer when applying this transformation.

Some teams applied this trick, but most teams did not.
In Section \ref{sec:results}, we therefore show three result tables: the results for the official submissions (with some teams using the trick, and most not), the results where the trick is used by no one, and the results where the trick is used by everyone.
As discussed in Section \ref{sec:discussion}, in retrospect, it would have been better to take ASD as the measure for determining the winners, since it is strongly related to ACC but does not benefit from the 2-5-trick.
However, the winning team is the same with respect to both ACC and ASD, and so is the runner-up.

The setup, datasets, rules, and measures, are also described in detail on the website of the triple scoring task: \url{http://www.wsdm-cup-2017.org/triple-scoring.html}.

\subsection{Related publications}

The WSDM Cup 2017 offered two tasks: vandalism detection and triple scoring.
The two tasks were completely independent, and only loosely related in the sense that both address a fundamental challenge when working with a very large knowledge base.

In \cite{DBLP:conf/wsdm/HeindorfPBBH17}, a brief overview of both tasks was provided.
The deadline for that overview paper was before the submission deadline for both tasks, hence it contains no information on the participating teams and the main results.

The triple scoring problem was introduced in \cite{DBLP:conf/sigir/BastBH15}.
The paper describes how the ground truth is obtained (via crowd-sourcing), and it presents several approaches (old and new) for solving the problem as well as an extensive evaluation.
A survey of basic and advanced techniques for entity search, and more generally semantic search, is provided in \cite{DBLP:journals/ftir/BastBH16}.
Both papers were known to the participants.

The participating teams developed several new ideas for solving the problem.
13 teams submitted a notebook paper describing their approach, see Table \ref{table:teams}.

\section{Participating teams}

Overall, 52 teams registered on the website, of which 33 also registered on TIRA (the platform we used for submitting runs in a reproducible fashion).
Eventually, 21 teams made a valid submission before the deadline.
The names and affiliations of these 21 teams are listed in Table \ref{table:teams}.
The team names were randomly assigned by the task organizers (us) from a list of healthy vegetables.

\def\Hvsep{\\[2mm]\hline\\[-3mm]}
\def\Hhsep{\hspace{5mm}}

\begin{table}[h]
{\renewcommand{\baselinestretch}{1.3}\normalsize
\begin{tabular}{@{\hspace{2mm}}l@{\hspace{5mm}}l@{\hspace{3mm}}l@{\hspace{2mm}}}
{\bf Team} & {\bf Affiliation} & {\bf \makebox[5mm][l]{Country}} \Hvsep
Bokchoy \cite{bokchoy}       & Chinese Academy of Sciences     & CN \\ %  1
Bologi                       & University of Illinois (UIUC)   & US \\ %  2
Cabbage \cite{cabbage}       & Negev + Tel Aviv University     & IL \\ %  3
Catsear \cite{catsear}       & University of Leipzig           & DE \\ %  4
Cauliflower                  & Yahoo! Japan                    & JP \\ %  5
Celosia \cite{celosia}       & IIIT Hyderabad + Microsoft      & IN \\ %  6
Chaya                        & KAIST                           & KR \\ %  7
Chickweed                    & Austral University              & AR \\ %  8
Chicory \cite{chicory}       & Radboud University + Spinque    & NL \\ %  9
Cress \cite{cress}           & NTNU Trondheim + U Stavanger    & NO \\ % 10
Endive                       &                                 & IR \\ % 11
Fiddlehead \cite{fiddlehead} & Fuji Xerox                      & JP \\ % 12
Gailan \cite{gailan}         & Trinity College                 & IE \\ % 13
Goosefoot \cite{goosefoot}   & Sofia University + QCRI         & BG+QA \\ % 14
Kale                         & University of Mannheim          & DE \\ % 15
Lettuce \cite{lettuce}       & Studio Ousia + Nara Institute   & JP \\ % 16
Pigweed \cite{pigweed}       & Yahoo! Japan                    & JP \\ % 17
Radicchio \cite{radicchio}   & University of Illinois (UIUC)   & US \\ % 18
Rapini                       &                                 & US \\ % 19
Samphire \cite{samphire}     & Indiana University Bloomington  & US \\ % 20
Yarrow                       & University of Illinois (UIUC)   & US \\ % 21
\end{tabular}}
\vspace{-2mm}
\caption{\label{table:teams}The 21 teams (in alphabetical order) who submitted a valid run, with their affiliations and publication if available.}
\end{table}

\section{Main Results}\label{sec:results}

Table \ref{table:results-official} shows the top 10-results of the official submissions with respect to each of the three quality measures explained in Section \ref{sec:data}.
As explained in Section \ref{sec:data}, if a team submitted several runs, the last run that was submitted before the deadline was taken.

\def\Htop#1{{#1}}
\def\Hu{\hspace{0.5mm}$\uparrow$}
\def\Hd{\hspace{0.5mm}$\downarrow$}
\def\Hvspacebeforecaption{\vspace{-3mm}}
\def\Hvspaceaftercaption{\vspace{3mm}}

\begin{table*}
\begin{center}
{\renewcommand{\baselinestretch}{1.3}\normalsize
\begin{tabular}{rll}
{\bf \#} & {\bf Team} & {\bf ACC}\Hvsep
\Htop{1.} & \Htop{Bokchoy*  } & \Htop{0.868} \\
\Htop{2.} & \Htop{Lettuce*  } & \Htop{0.823} \\
\Htop{3.} & \Htop{Radicchio } & \Htop{0.797} \\
 4. & Catsear       & 0.796 \\
 5. & Samphire*     & 0.780 \\
 6. & Cress         & 0.779 \\
 7. & Chickweed     & 0.772 \\
 8. & Cauliflower*  & 0.752 \\
 9. & Goosefoot     & 0.747 \\
10. & Cabbage       & 0.737 \\[2mm]
14. & Baseline      & 0.721
\end{tabular}
\Hhsep
\begin{tabular}{rll}
{\bf \#} & {\bf Team} & {\bf ASD}\Hvsep
 1. & Cress        & 1.613 \\
 2. & Bokchoy*     & 1.630 \\
 3. & Radicchio    & 1.692 \\
 4. & Fiddlehead   & 1.704 \\
 5. & Cabbage      & 1.735 \\
 6. & Lettuce*     & 1.762 \\
 7. & Goosefoot    & 1.776 \\
 8. & Chaya        & 1.811 \\
 9. & Gailan       & 1.837 \\
10. & Kale         & 1.855 \\[2mm]
21. & Baseline     & 2.070
\end{tabular}
\Hhsep
\begin{tabular}{rll}
{\bf \#} & {\bf Team} & {\bf TAU}\Hvsep
 1. & Goosefoot    & 0.314 \\
 2. & Cress        & 0.321 \\
 3. & Bokchoy*     & 0.327 \\
 4. & Chaya        & 0.337 \\
 5. & Chicory      & 0.353 \\
 6. & Cabbage      & 0.354 \\
 7. & Lettuce*     & 0.362 \\
 8. & Kale         & 0.363 \\
 9. & Gailan       & 0.370 \\
10. & Chickweed    & 0.392 \\[2mm]
20. & Baseline     & 0.460
\end{tabular}}
\end{center}
\Hvspacebeforecaption
\caption{\label{table:results-official}Top-10 results of the official submissions and the baseline. The teams marked * rounded all scores to the range 2..5 (the 2-5-trick described in Section \ref{sec:data}) for their official submission.}
\Hvspaceaftercaption
\end{table*}

\begin{table*}
\begin{center}
{\renewcommand{\baselinestretch}{1.3}\normalsize
\begin{tabular}{rll}
{\bf \#} & {\bf Team} & {\bf ACC}\Hvsep
 1. & Bokchoy       & 0.818 \\
 2. & Radicchio\Hu  & 0.797 \\
 3. & Catsear\Hu    & 0.796 \\
 4. & Cress\Hu      & 0.779 \\
 5. & Lettuce\Hd    & 0.772 \\
 6. & Chickweed\Hu  & 0.772 \\
 7. & Goosefoot\Hu  & 0.746 \\
 8. & Cabbage\Hu    & 0.737 \\
 9. & Pigweed\Hu    & 0.737 \\
10. & Fiddlehead\Hu & 0.728 \\[2mm]
11. & Baseline\Hu   & 0.721
\end{tabular}
\Hhsep
\begin{tabular}{rll}
{\bf \#} & {\bf Team} & {\bf ASD}\Hvsep
 1. & Bokchoy\Hu    & 1.501 \\
 2. & Lettuce\Hu    & 1.594 \\
 3. & Cress\Hd      & 1.613 \\
 4. & Radicchio\Hd  & 1.692 \\
 5. & Fiddlehead\Hd & 1.704 \\
 6. & Cabbage\Hd    & 1.735 \\
 7. & Goosefoot     & 1.776 \\
 8. & Chaya         & 1.811 \\
 9. & Gailan        & 1.837 \\
10. & Kale          & 1.855 \\[2mm]
20. & Baseline\Hu   & 2.070
\end{tabular}
\Hhsep
\begin{tabular}{rll}
{\bf \#} & {\bf Team} & {\bf TAU}\Hvsep
 1. & Lettuce\Hu    & 0.294 \\
 2. & Goosefoot\Hd  & 0.314 \\
 3. & Bokchoy       & 0.316 \\
 4. & Cress\Hd      & 0.321 \\
 5. & Chaya\Hd      & 0.337 \\
 6. & Cabbage       & 0.354 \\
 7. & Kale\Hu       & 0.363 \\
 8. & Gailan\Hu     & 0.370 \\
 9. & Chickweed\Hu  & 0.392 \\
10. & Fiddlehead\Hu & 0.395 \\[2mm]
20. & Baseline      & 0.460
\end{tabular}}
\end{center}
\Hvspacebeforecaption
\caption{\label{table:results-fullrange}Top-10 results of the submissions with scores in the full 0..7 range, as well as the (unchanged) baseline. The arrows indicate if the ranking of a team improved or worsened compared to the respective column in Table \ref{table:results-official}}
\Hvspaceaftercaption
\end{table*}

\begin{table*}
\begin{center}
{\renewcommand{\baselinestretch}{1.3}\normalsize
\begin{tabular}{rll}
{\bf \#} & {\bf Team} & {\bf ACC}\Hvsep
 1. & Bokchoy       & 0.868 \\
 2. & Lettuce       & 0.823 \\
 3. & Radicchio     & 0.814 \\
 4. & Cabbage\Hu    & 0.813 \\
 5. & Gossefoot\Hu  & 0.804 \\
 6. & Cress         & 0.797 \\
 7. & Catsear\Hd    & 0.796 \\
 8. & Chaya\Hu      & 0.786 \\
 9. & Filddehead\Hu & 0.782 \\
10. & Chickweed\Hd  & 0.782 \\[2mm]
18. & Baseline\Hd   & 0.721
\end{tabular}
\Hhsep
\begin{tabular}{rll}
{\bf \#} & {\bf Team} & {\bf ASD}\Hvsep
 1. & Bokchoy\Hu    & 1.630 \\
 2. & Cabbage\Hu    & 1.735 \\
 3. & Radicchio     & 1.754 \\
 4. & Goosefoot\Hu  & 1.755 \\
 5. & Lettuce\Hu    & 1.762 \\
 6. & Cress\Hd      & 1.768 \\
 7. & Fiddlehead\Hd & 1.820 \\
 8. & Bologi\Hu     & 1.823 \\
 9. & Chaya\Hd      & 1.827 \\
10. & Catsear\Hu    & 1.859 \\[2mm]
21. & Baseline      & 2.070
\end{tabular}
\Hhsep
\begin{tabular}{rll}
{\bf \#} & {\bf Team} & {\bf TAU}\Hvsep
 1. & Bokchoy\Hu   & 0.327 \\
 2. & Chaya\Hu     & 0.332 \\
 3. & Cress\Hd     & 0.335 \\
 4. & Goosefoot\Hd & 0.343 \\
 5. & Cabbage\Hu   & 0.359 \\
 6. & Lettuce\Hu   & 0.362 \\
 7. & Chicory\Hd   & 0.368 \\
 8. & Bologi\Hu    & 0.385 \\
 9. & Chickweed\Hu & 0.386 \\
10. & Kale\Hd      & 0.389 \\[2mm]
20. & Baseline     & 0.460
\end{tabular}}
\end{center}
\Hvspacebeforecaption
\caption{\label{table:results-25range}Top-10 results of the submissions with scores rounded to the range 2..5, as well as the (unchanged) baseline. The arrows indicate if the ranking of a team improved or worsened compared to the respective column in Table \ref{table:results-official}.}
\end{table*}

Included are the results for a very simple \emph{baseline} method, which simply assigns score 5 to each triple.
This is a reasonable baseline, because this is the one fixed score that gives the best result on the training set with respect to ACC.
The reason is that the majority of triples are relevant and that with respect to ACC the score 5 is the best choice for a relevant triple, since it counts as accurate when the score in the ground truth is $3$, $4$, $5$, $6$, or $7$.

Since the three official winners of the competition were selected according to ACC (first column of Table \ref{table:results-official}), several teams exploited the 2-5-trick explained in Section \ref{sec:data}: computing only scores from the range 2..5. As explained above, this can (and usually does) harm ASD and TAU, but can only improve ACC and never make it worse. The teams who exploited this trick are marked with a * in Table \ref{table:results-official}.

As some teams exploited this trick to improve their chances of winning the prize money (which was a completely reasonable thing to do) yet most teams did not, the results from these two groups are somewhat hard to compare. We therefore also provide an evaluation of two variants of the submitted runs.

Table \ref{table:results-fullrange} shows the results, when all teams use the full score range. For this table, the teams marked * in Table \ref{table:results-official} were individually asked (outside of the competition) to run a variant of their method that uses the full score range (thus likely making their ACC scores worse, but likely improving their ASD and TAU scores).

Table \ref{table:results-25range} shows the results, when all the scores from the official submissions were truncated to the range 2..5.
This transformation was trivial to apply (and did not change anything for the teams marked * in Table \ref{table:results-official}).
%, who submitted runs with scores in that range in the first place).

%\vspace{-3mm}
\section{Discussion}\label{sec:discussion}

% We make the following observations:

With respect to ACC, almost all teams that did not use the 2-5-trick benefit quite significantly from the truncation to the range 2..5. For the top-10 teams from the first column (ACC) of Table \ref{table:results-official}, the improvement ranges from 0 (Catsear) to 0.076 (Cabbage).

Team Bokchoy, which is the winner with respect to ACC, is also in the top-3 with respect to ASD and TAU, despite using the 2-5-trick. With scores in the full range for all teams, team Bokchoy also wins with respect to ASD, and with scores in the range 2..5 for all teams, team Bokchoy also wins with respect to TAU. This demonstrates the particular strength of their approach, which is briefly described in Section \ref{sec:methods} below.

Performance with respect to ACC and ASD is strongly related for all teams. This is intuitive, because both measures require a good estimate of the score from the ground truth.

Performance with respect to TAU is only weakly related to performance with respect to ACC or ASD. For example, none of the teams on places 3, 4, and 5 with respect to ACC are in the top-10 with respect to TAU. Indeed, ranking is a relatively simpler task than estimating the score. For example, when a subject has only two objects, finding the right ranking is a binary classification problem, whereas estimating the score has more degrees of freedom.

For future evaluations of this task, ASD should be preferred to ACC, because ASD does not benefit from the 2-5-trick, but is otherwise strongly related to ACC. TAU is a measure of independent interest. If only one measure should be singled out, ASD is the measure of choice, because minimizing ASD is harder than minimizing TAU. However, it should be noted that some applications may only require a ranking and not the exact scores, in which case TAU is a perfectly reasonable measure.

The simple baseline beats one third of the teams with respect to ACC. This is simply due to the fact, that many teams did not use the 2-5-trick (and thus did not optimize strictly for ACC). With respect to ASD and TAU, the baseline ranks poorly, as it should.

%\newpage
\section{Methods employed by the participants}\label{sec:methods}

All 21 teams who submitted a valid run made use of the Wikipedia sentences provided by the organizers; see Section \ref{sec:data}.
Many (but not all) teams used variants of the ideas presented in \cite{DBLP:conf/sigir/BastBH15}.
For details, we refer the interested reader to the various notebook papers referenced in Table \ref{table:teams}.

We briefly summarize the approach of team Bokchoy \cite{bokchoy}, who did not only win the competition, but were among the top-3 with respect to all three quality measures (ACC, ASD, TAU) and in all three variants of our evaluation (Tables \ref{table:results-official}, \ref{table:results-fullrange}, and \ref{table:results-25range}). Their approach has two main components.

The first component is an ensemble learner using four independent approaches for computing triple scores. Three of these approaches were taken from the original paper \cite{DBLP:conf/sigir/BastBH15}, all using the Wikipedia sentences. The fourth approach makes use of Freebase paths between the subject and object of the triple to be scored. These paths are used as features for a binary classifier that decides whether the triple should get a high score or a low score. For example, the path \emph{Johnny Depp born-in Kentucky located-in USA} is a strong indication that the triple \emph{Johnny Depp nationality USA} should get a high score.

The second component can promote or demote scores based on so-called trigger words for a type (which in this task was either a nationality or a profession).
These are words that are semantically very strongly related to the type.
For example, trigger words for the German nationality are \emph{German} and \emph{Germany}.
Trigger words were compiled from a variety of sources, including a list of adjectival forms, synonyms and hyponyms from WordNet, and some manually compiled words.
The trigger words were then employed as follows to modify the score $s$ output by the ensemble learner, using the first paragraph of the Wikipedia article of the subject of the triple in question.
If \emph{the first sentence} of that paragraph contains at least one of the trigger words, promote the score to $\max\{5, s\}$.
If \emph{the whole paragraph} contains none of the trigger words, demote the score to $\min\{2, s\}$.

The intuition behind this second component is that a mention of a trigger word in the first sentence of a Wikipedia article is a strong signal for a high relevance of the respective information.
Similarly, if no trigger word is mentioned in the whole first paragraph (which in case of a person entity is usually a synopsis of the main aspects of that person's life), this is a strong signal for a low relevance of the respective information.
This simple post-processing of the scores improves the already good quality of the ensemble scorer significantly.

\section{Conclusion}

We have presented and discussed the results of the triple scoring task of the WSDM Cup 2017.
Triple scoring is a basic ingredient in entity ranking, that is, for searches that return a list of entities.
The task has attracted considerable interest from all over world, with 52 initial registrations and 21 teams submitting a valid run before the deadline.
A summary of the task and all the relevant data is freely available from the task's website: \url{http://www.wsdm-cup-2017.org/triple-scoring.html}\footnote{Should this URL ever cease to function, search for the keywords \emph{triple scoring wsdm cup 2017} and you should find another page with the same content.}.
We hope that the task and this overview and the free availability of the data spark further research on this important and interesting problem.

%%%%%%%%%%%%%%%%%%%%%%%%%%%%%%%%%%%%%%%%%%%%%%%%%%%%%%%%%%%%%%%%%%%%%%%%
\newpage
{\raggedright
\bibliography{wsdm-cup-triple-scoring}}
\end{document}